\documentclass[twocolumn,preprintnumbers,amsmath,amssymb]{revtex4}

\usepackage{graphicx}
\usepackage{dcolumn}
\usepackage{bm}
\usepackage{hyperref}
\usepackage{amsmath}
\usepackage{epstopdf}
\usepackage{color}
\usepackage{verbatim}
\usepackage[normalem]{ulem}
\definecolor{LightGrey}{RGB}{211, 211, 211}

\begin{document}

\title{Strain-induced InGaAs/InAlAs superlattices for terahertz radiation}

\author{\bf D S Ponomarev$^{1,2,3}$, A E Yachmenev$^{1,2}$, S S Pushkarev$^1$, R A Khabibullin$^{1,2,3}$, M M Grekhov$^4$, A~Gorodetsky$^{5,6}$, K I~Zaytsev$^{2,7,8}$, D I~Khusyainov$^9$, A M Buryakov$^9$, E D Mishina$^9$
 }

\affiliation
{$^1$Institute of Ultra High Frequency Semiconductor Electronics RAS, Moscow 117105, Russia\\
$^2$Prokhorov General Physics Institute of RAS, Moscow 119991, Russia\\
$^3$Center for Photonics and 2D Materials,
Moscow Institute of Physics and Technology,
Dolgoprudny 141700, Russia\\
$^4$National Research Nuclear University MEPhI, Moscow 115409, Russia\\
$^5$ITMO University, St. Petersburg 197101, Russia\\
$^6$Department of Chemistry, Imperial College London, London SW7 2AZ, UK\\
$^7$Bauman Moscow State Technical University, Moscow 105005, Russia\\
$^8$Sechenov First Moscow State Medical University, Moscow 119991, Russia\\
$^9$Russian Technological University, Moscow 119454, Russia\\
}

\begin{abstract}
We study the influence of residual strain in InGaAs/InAlAs superlattice (SL) on the ultrafast photocarrier dynamics under femtosecond laser excitation. We propose and fabricate a novel-designed strained InGaAs/InAlAs SL 
which allows us to obtain an ultrashort photocarrier relaxation time of $\tau\sim$ 1.5 ps without Be-doping of the InGaAs photoconductor. We assume two dominant mechanisms to be responsible for a sharp reduction of $\tau$: photocarriers trapping by defect levels in InAlAs barriers and increased photocarriers scattering at InGaAs/InAlAs 
interface roughness due to residual strain in the SL. 
The THz time-domain spectroscopic measurements reveal an increase in both emitted THz waveform and spectrum amplitudes with an increase of residual strain in SL. The results might be of considerable interest for accommodating the needs of THz pulsed spectroscopy and imaging in fundamental and applied branches of THz science and technology.

\end{abstract}

\maketitle

\section{Introduction}

Terahertz (THz) frequency range of the electromagnetic spectrum is of considerable interest due to its wide applications in security systems, biology, medicine and material science. To date the research and development of compact, broadband and coherent sub-THz and THz emitters and detectors are among the hottest issues \cite{Opt.Engineering.REVIEW.Burford.2017, Xurong, Ponomarev_HEB, Polischuk, Chernomyrdin, Lepeshov.2017.Enhancement, Gorodetsky1, LepeshovSCIREP, Nandi2018, Fadeev}.
The most commonly used state-of-art THz emitters are based on photoconductive antennas (PCAs) where photocarries are accelerated by applied external bias \cite{Castro-Camus.PCAreview.2016, Jepsen:96, 11,APE.Krotkus.GaAsBi.2012}. Large external biases are essential for increasing a peak THz power leading to restrictions on photoconductors, namely on its carrier relaxation times and resistance.
The ternary compound InGaAs is a promising photoconductor since it operates with long-wave laser pump \mbox{1.0$-$1.55 $\mu$m} emitted by fiber lasers that allows for fabricating low-cost and compact THz emitters for THz spectroscopy and imaging \cite{Brahm:14},\cite{Dietz:11}. When an ultrafast laser pulse with photon energy above the semiconductor bandgap strikes InGaAs, electron-hole pairs are generated at sub-surface of the semiconductor. These pairs are rapidly accelerated by the built-in electric field resulting in a radiative dipole parallel to surface normal \cite{Auston.1983.APL},\cite{Hwang:07}. Also THz emission may occur as a result of the boundary conditions on the carrier transport within a narrow-band semiconductor as a result of the photo-Dember (PD) effect \cite{Klatt:10},\cite{Reklaitis}. 
Recently THz emission has been observed in InGaAs nanowire arrays \cite{Beleckaite} as well as in iron-doped InGaAs \cite{Globisch}.

Nevertheless InGaAs provides relatively high carrier relaxation times and low resistivity and thus suffers from high dark currents which are detrimental for biased PCAs. In order to improve both parameters different approaches have been proposed, among them are Be-doped low-temperature grown InGaAs \cite{Globisch2, OHTSUKA1991460, Koumetz}, ion-implantation in InGaAs \cite{Suzuki},\cite{Chimot}, incorporating of ErAs islands into InGaAs/InAlAs quantum wells \cite{Yang2016, Sukhotin}, substrates containing self-assembled InAs quantum dots in GaAs matrix~\cite{Leyman2016,Fedorova2016a}, etc.
An efficient approach for reaching a trade-off between a carrier relaxation time and a resistivity is in using the InGaAs/InAlAs superlattices (SLs). In Ref.\cite{Ospald} authors proposed a SL consisting of the self-assembled ErAs layers separated by the InGaAs photoconductors. The deposition of Er-atoms onto the InGaAs surface allows for formation of ErAs islands. On the one hand, the incorporation of ErAs into the InGaAs crystal lattice results in a shift of the Fermi level to the bottom of the InGaAs conduction band, accompanied by an increase in the concentration of photocarriers in the photoconductor and with a corresponding decrease in a resistivity of the layer; on the other hand, with Be-doping InGaAs, by a proper choice of acceptors density, the Fermi level is pinned approximately in the middle of the InGaAs band gap. With a SL consisting of 30$-$70 ErAs:InGaAs periods, a short carrier relaxation time, $\tau\sim$ 3.6 ps, was attained. Sukhotin et al.\cite{Sukhotin3} additionally used a smoothing InAlAs layer, which resulted in even a larger decrease in relaxation time up to $\tau\sim$~2~ps.
Lately Roehle et al.\cite{Roehle:10} proposed a SL, in which InGaAs photoconductive layers were sandwiched between InAlAs barrier layers. The reduced growth temperature of barrier layers provides a large number of deep levels that serve as traps for photocarriers \cite{Dietz:11, Dietz64}. When tunneling between InGaAs layers, they are captured in InAlAs barriers by traps that serve as recombination centers $-$ this results in an increase of InGaAs resistance. When using a SL consisting of 100 periods of Be-doped InGaAs photoconductive layers, with the total active region thickness of 1$\mu$m, the record short photocarriers relaxation time of $\tau\sim$~0.8~ps was obtained.

Recently, we demonstrated that epitaxial stresses in the InGaAs photoconductor allow for reduction of photocarrier relaxation times up to 20\%
 in comparison to non-strained InGaAs due to a formation of recombination centers in the band gap of a photoconductor \cite{Khusyainov2017, KhusyainovPJTF}. Lately we fabricated a 30-period SL slightly analogous to that in Ref.\cite{Roehle:10} but grown on GaAs wafer by means of a metamorphic buffer (MB) which showed the photocarrier relaxation time of \mbox{$\tau\sim$ 3.4 ps} without Be-doping the InGaAs photoconductor \cite{Ponomarev2018}. Since the introduction of strain in epitaxial layers leads to increasing of an interface roughness\cite{Strain1,Strain2}, in present paper, we combine these two approaches and present a novel-designed strained \mbox{30-period InGaAs/InAlAs SL} that allows us to obtain an ultrashort photocarrier relaxation time of $\tau\sim$~1.5~ps at the pump power of 1~mW. We assume the two dominant mechanisms being responsible for a sharp reduction of $\tau$ that are: photocarriers trapping by defect levels in InAlAs barriers and its increased scattering at the InGaAs/InAlAs interface roughness due to residual strain in the SL.


\section{Modeling of band structure}
In order to calculate the electronic states in the
$\text{In}_{0.53}\text{Ga}_{0.47}\text{As/In}_{0.52}\text{Al}_{0.48}\text{As}$ SL and to estimate the
overlapping of electron wave functions (WFs) with the
trapping states in $\text{In}_{0.52}\text{Al}_{0.48}\text{As}$ barriers, we solved the
Schr\"{o}dinger equation in the effective mass approximation. The parameters of the semiconductors
and alloys were taken from Ref.\cite{Vurgaftman}. The offsets of the conduction
band at the interface between the InGaAs
and InAlAs layers in relation to the indium content were calculated by the linear interpolation
procedures described in Ref.\cite{Krijn}, using the band offsets for
the GaAs, InAs, and AlAs binary semiconductors
from Ref.\cite{Wei}. We considered the non-strained $\text{In}_{0.53}\text{Ga}_{0.47}\text{As/In}_{0.52}\text{Al}_{0.48}\text{As}$ SL. The detailed simulation procedure is given in Ref.\cite{Ponomarev2018}, while the results of simulation are depicted in Fig.1.

\begin{figure}[h!]
  \centering
    \includegraphics[width=1.0\columnwidth]{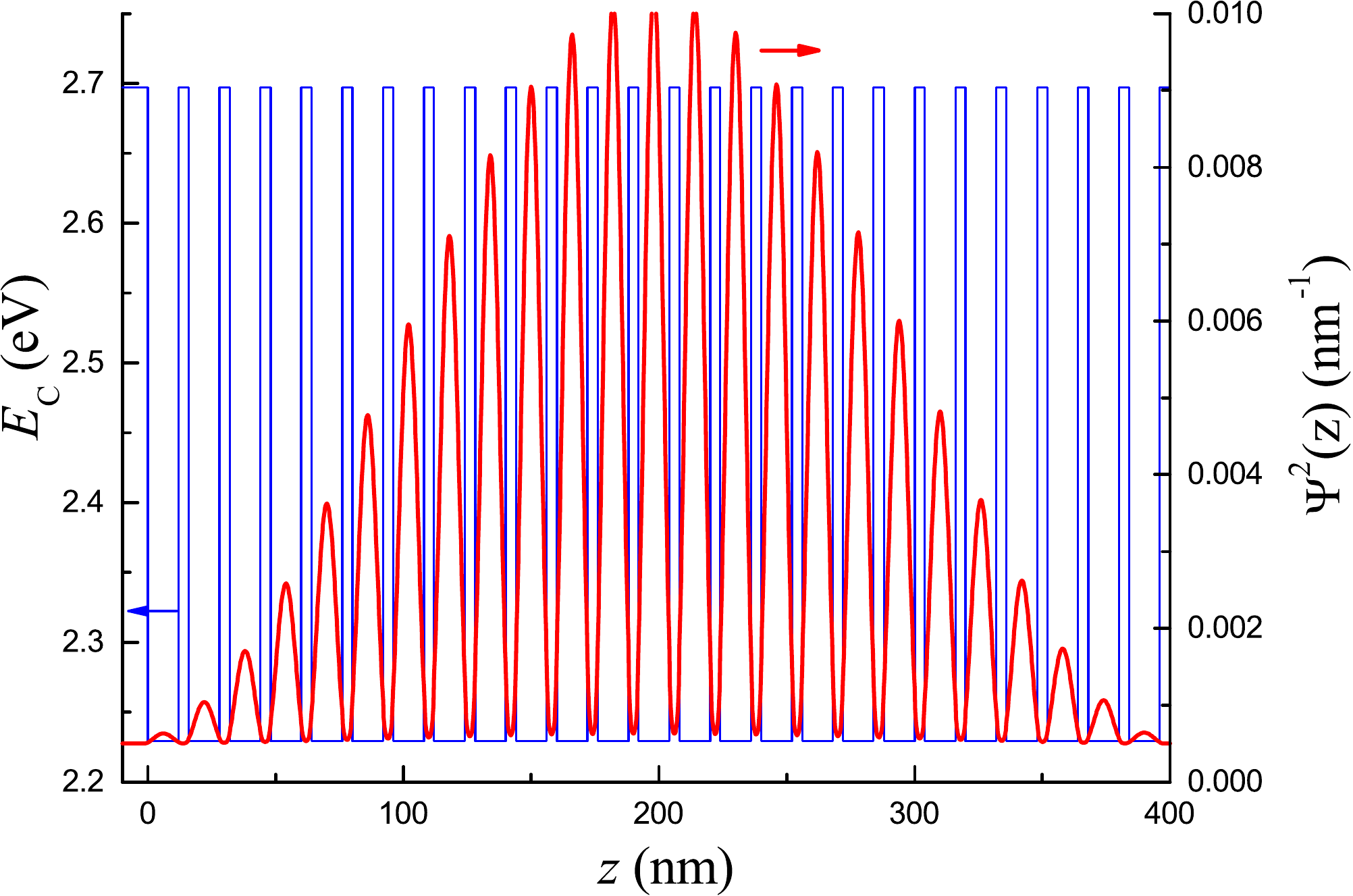}
    \caption{\small{Conduction-band profile ($E_{c}$) and the eigen wave function of the ground
electron state $\Psi^{2}$(z) in the non-strained $\text{In}_{0.53}\text{Ga}_{0.47}\text{As/In}_{0.52}\text{Al}_{0.48}\text{As}$ superlattice with
the $\text{In}_{0.53}\text{Ga}_{0.47}\text{As}$ and $\text{In}_{0.52}\text{Al}_{0.48}\text{As}$ thicknesses equal to 12~nm and 4~nm, respectively; $z$ is a growth direction.}}
    \label{Simulation}
\end{figure}

First, we determined the
decay parameter of electron WFs ($L_b$) in the
$\text{In}_{0.52}\text{Al}_{0.48}\text{As}$ barrier layer in the case of a single quantum
well (QW). To do this, we solved the quantum-mechanical
problem of an electron in a rectangular
$\text{In}_{0.53}\text{Ga}_{0.47}\text{As/In}_{0.52}\text{Al}_{0.48}\text{As}$ QW with the width of $L$ =
12 nm. In this problem, the position of the lower
energy level $E_1$ is determined by:
\begin{equation}
\frac{m_w}{m_b}\frac{k_w}{k_b}\tan\left(\frac{k_w L}{2}\right)=1.
\label{Eq1}
\end{equation}
where $m_{w}$, $m_{b}$ are the electron effective masses in QW and barriers respectively, $L$ is the width of QW, \mbox{$k_{w}=\sqrt{\frac{2m_wE_1}{\hbar^{2}}}$}, \mbox{$k_{b}=\sqrt{\frac{2m_b(U_{0}-E_{1})}{\hbar^{2}}}$} and $U_{0}$ is the depth of QW. The parameter $k_{b}$ describes the decay length of
the electron WF $\Psi$(z) in the barrier, since the WF in
this region is described by the exponentially decaying
function $\Psi(z) \sim e^{(-k_bz)}$, where $z$ is a growth direction. Equation (1) was solved numerically by searching one-dimensional roots by bisecting
the segment, then using the previously determined
energy of the ground state, we calculated the decay
parameter of the electron WF in the barrier \mbox{$L_{b} = 1/(k_{b})$}.

Second, we calculated the
penetration depth of the electron WF into the
$\text{In}_{0.52}\text{Al}_{0.48}\text{As}$  barrier in the SL. It describes the region where electrons
interact with defects and trapping states in
$\text{In}_{0.52}\text{Al}_{0.48}\text{As}$  barrier layers. Since in the SL under consideration
electron WFs of neighboring QWs overlap, it leads to a formation of SL
states delocalized within several QWs.
In this case, the electron WF in barriers is not
described by a simple analytical function, as in the
case of a single rectangular QW. Therefore, to assess
the distribution of the electron density in the
$\text{In}_{0.53}\text{Ga}_{0.47}\text{As/In}_{0.52}\text{Al}_{0.48}\text{As}$ SL, we solved the
Schr\"{o}dinger equation numerically. \mbox{Figure 1} demonstrates the
results of numerical simulation of the band diagram
and the WF of the electron ground state for the
$\text{In}_{0.53}\text{Ga}_{0.47}\text{As/In}_{0.52}\text{Al}_{0.48}\text{As}$ SL with the $\text{In}_{0.53}\text{Ga}_{0.47}\text{As}$ and $\text{In}_{0.52}\text{Al}_{0.48}\text{As}$
thicknesses of 12~nm and 4~nm, respectively. As it can be seen in Fig.~\ref{Simulation}, the electron WF is
delocalized over the entire thickness of the SL while the electron WF in the $\text{In}_{0.52}\text{Al}_{0.48}\text{As}$ barriers exponentially
decreases along the direction from the boundaries of
the barriers to their centers. The calculation shows
that, for the SL with thick barriers (12 nm), the WF
amplitude at the barrier centers is nearly zero. It is
obvious that a barrier can be thought of as being thick
at $L \gg 2L_{b}$. For thin barriers with $L \sim 2L_{b}$ (see Fig. 1), the
electron WF strongly overlaps the barriers and is zero
nowhere except at the SL boundaries. According to the simulation results we chose the optimal parameters of the SL which were then used when fabricating the experimental samples (see Section III).
The results are in good agreement with the data, previously reported in Refs.\cite{Dietz2015-4, Dietz64}.

\section{Results and discussion}
\subsection{Samples fabrication}
The samples of the SLs were grown by molecular-beam epitaxy (MBE) on Riber 32 system with solid-state sources on the semi-insulating (100) GaAs wafers. A step-graded MB was used for matching the lattice parameters between the substrate and the SL, and consisted of 5 InAlAs layers of 0.15$\mu$m thickness with an increase in indium mole fraction by 10\%
in each consequent layer. The first sample (287) comprises of 30 period non-strained $\text{In}_{0.53}\text{Ga}_{0.47}\text{As/In}_{0.52}\text{Al}_{0.48}\text{As}$ SL with layers thicknesses of 12 nm and 4 nm respectively, according to the results of numerical calculations. Its schematic layout is given in Ref.~\cite{Ponomarev2018}. The second sample (288) had the analogous SL but its $\text{In}_{0.52}\text{Al}_{0.48}\text{As}$ barrier layers were lattice-mismatched with the $\text{In}_{0.53}\text{Ga}_{0.47}\text{As}$ photoconductive layers by decreasing the indium mole fraction to $\text{In}_{0.38}\text{Al}_{0.62}\text{As}$ which resulted in appearance of residual strain in the SL. The total thicknesses of the $\text{In}_{0.53}\text{Ga}_{0.47}\text{As}$ photoconductive layers and the $\text{In}_{0.52}\text{Al}_{0.48}\text{As}$, $\text{In}_{0.38}\text{Al}_{0.62}\text{As}$ barrier layers were 0.36$\mu$m and 0.12$\mu$m respectively, the growth temperature was equal to \mbox{$T$ = 400 $^{\circ} C$}, for all grown samples.



\subsection{Structural analysis. Residual strain estimation}

Structural analysis of the samples under consideration was performed by high-resolution double-crystal X-ray diffractometry. Diffraction reflection curves (DRCs) in the $\theta/2\theta$ scanning mode were measured by means of Ultima IV double-crystal X-ray diffractometer (XRD, Rigaku) using CuK$_{\alpha1}$ radiation with a wavelength of \mbox{$\lambda$ = 1.54056 ${\AA}$} and power of \mbox{0.9 kW}. For symmetric reflections, the (400) plane was chosen while for estimating a residual strain additional DRCs were measured using asymmetric reflections from the (411) plane. The measurement methodic is described in details in Ref.\cite{Ponomarev2017}.

The DRCs for the strained SL (288) is depicted in Fig.2 [for comparison we also attach the DRC for the non-strained sample (287)]. The labels I, II, III, IV, V refer
to the 5 steps (layers) of the MB, which are compressively strained. The lateral strain defined as \mbox{$\epsilon$ = ($a_{\text {lateral}}$ $-$ $a_{\text {relax}}$)/$a_{\text {relax}}$}, where $a_{\text {lateral}}$ and $a_{\text {relax}}$ are lateral and relaxed lattice parameters respectively, arises from the bottom toward the top of MB and equals to -0.0005, -0.0016, -0.0017, -0.0013, -0.0052 for the sample 288, assuming tetragonal distortion of the MB crystal lattice.
Because of a sufficient dislocation density caused by the MB the interferential oscillations from the SLs on DRCs are rather small and broadened. 
We ascertain the residual strain as -0.0052/+0.0038 (for $\text{In}_{0.53}\text{Ga}_{0.47}\text{As/In}_{0.38}\text{Al}_{0.62}\text{As}$), which demonstrates that the strained SL is subjected to alternating opposite (compressive and tensile) strains which are modulo rather close.


\begin{figure}[h]
  \centering
    \includegraphics[width=0.9\columnwidth]{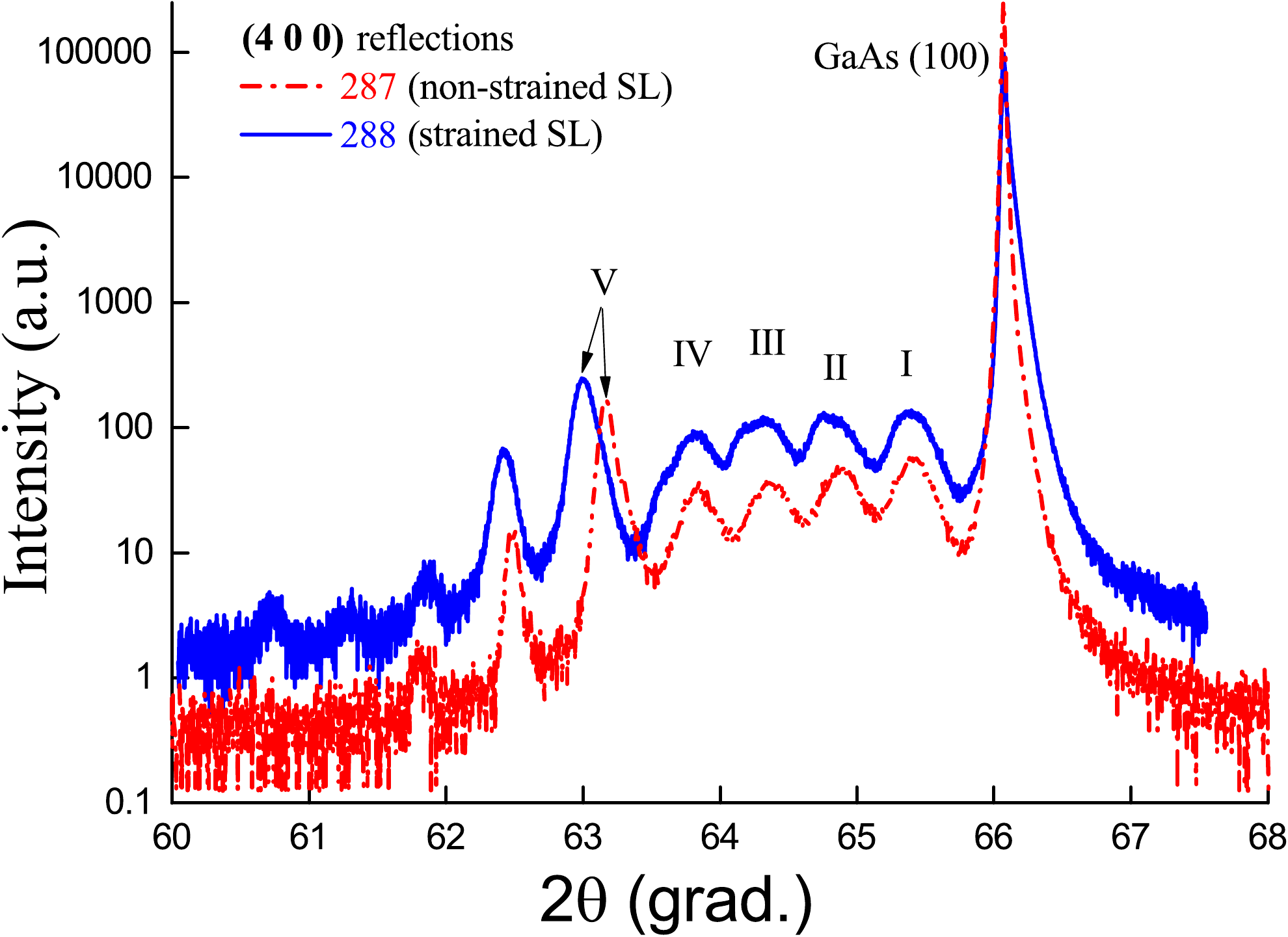}
    \caption{\small{Diffraction reflection curves for (400) symmetric reflections in the samples with the non-strained $\text{In}_{0.53}\text{Ga}_{0.47}\text{As/In}_{0.52}\text{Al}_{0.48}\text{As}$ (287) and
    strained $\text{In}_{0.53}\text{Ga}_{0.47}\text{As/In}_{0.38}\text{Al}_{0.62}\text{As}$ (288) superlattices. 
  The labels I, II, III, IV, V refer to 5 steps of the metamorphic buffer.}}
    \label{DRC}
\end{figure}

\subsection{Transmission electron microscopy characterization}
To characterize the grown samples of the SLs we carried out high resolution transmission electron microscopy (TEM). We limited the TEM investigation by the microscopy of the non-strained sample (287) to study the crystalline quality of high indium content epitaxial layers grown on GaAs substrate. Prior to electron microscopy, cross sections of the sample was prepared by standard methods. Mechanical thinning to 20$-$40$\mu$m was followed by Arion sputtering at an accelerating voltage of 5 keV in a Gatan 691 PIPS system (GATAN, United States) until a hole was formed. Final polishing was performed by ions with an energy reduced to 0.1 keV. The sample was analyzed in a TEM TITAN 80$-$300 (FEI, United States) with a corrector of probe spherical aberration in the bright- and dark-field modes. A high-angle annular dark-field (HAADF) detector of scattered electrons was used in the latter case with sample scanning. The accelerating voltage was 300 kV.

Figure 3 demonstrates a bright-field TEM-image of the non-strained $\text{In}_{0.53}\text{Ga}_{0.47}\text{As/In}_{0.52}\text{Al}_{0.48}\text{As}$SL (a) and its magnified microphotography (b) which showcase accurately 30 periods of $\text{In}_{0.53}\text{Ga}_{0.47}\text{As/In}_{0.52}\text{Al}_{0.48}\text{As}$ layers, labeled by 1$-$30 in (a), and confirm an absence of threading dislocations in active layers of the SL and its high crystalline quality.

\begin{figure}[h]
  \centering
    \includegraphics[width=1.0\columnwidth]{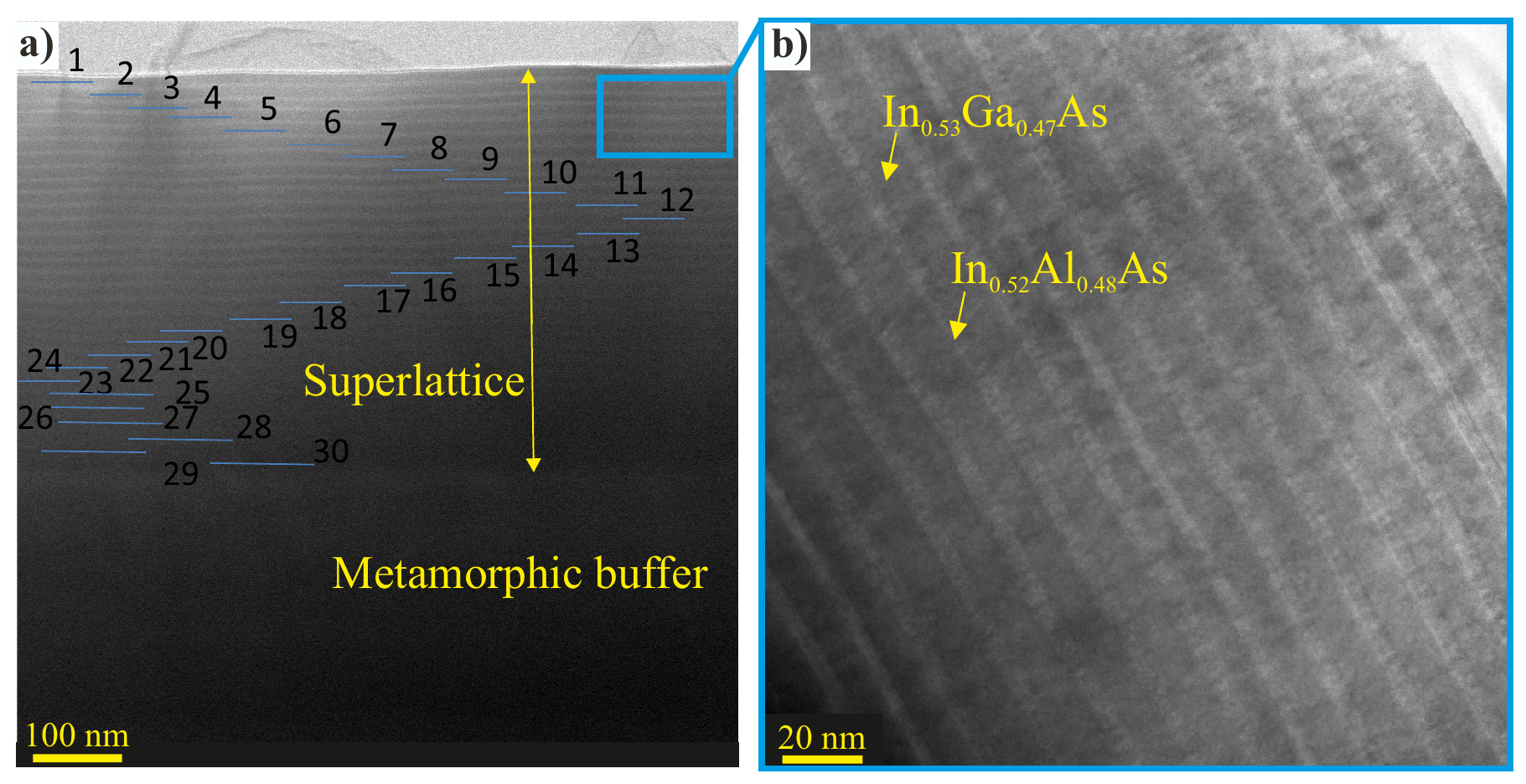}
    \caption{\small{Bright-field TEM-image of the non-strained $\text{In}_{0.53}\text{Ga}_{0.47}\text{As/In}_{0.52}\text{Al}_{0.48}\text{As}$ superlattice ($a$) and its magnified microphotography ($b$). The labels 1$-$30 refer to periods of the SL.}}
    \label{TEM}
\end{figure}

Figure 4 demonstrates a dark-field TEM-image of the non-strained sample. As clearly seen, the thicknesses of $\text{In}_{0.53}\text{Ga}_{0.47}\text{As}$ and $\text{In}_{0.52}\text{Al}_{0.48}\text{As}$ layers are equal to \mbox{12 nm} and \mbox{4 nm}, respectively. Since the $\text{In}_{0.52}\text{Al}_{0.48}\text{As}$ barrier layers were grown at the reduced temperature one can see three $\text{In}_{0.52}\text{Al}_{0.48}\text{As}$ sub-layers, approximately \mbox{1$-$2 nm} thick, which probably arise due to segregation of InAs and AlAs compounds and lead to formation of defect levels for trapping photocarriers when tunneling between the $\text{In}_{0.53}\text{Ga}_{0.47}\text{As}$ photoconductive layers. It should be noted that the experimental results are in good agreement with the results of the numerical calculations (see Section II).

\begin{figure}[h]
  \centering
    \includegraphics[width=0.7\columnwidth]{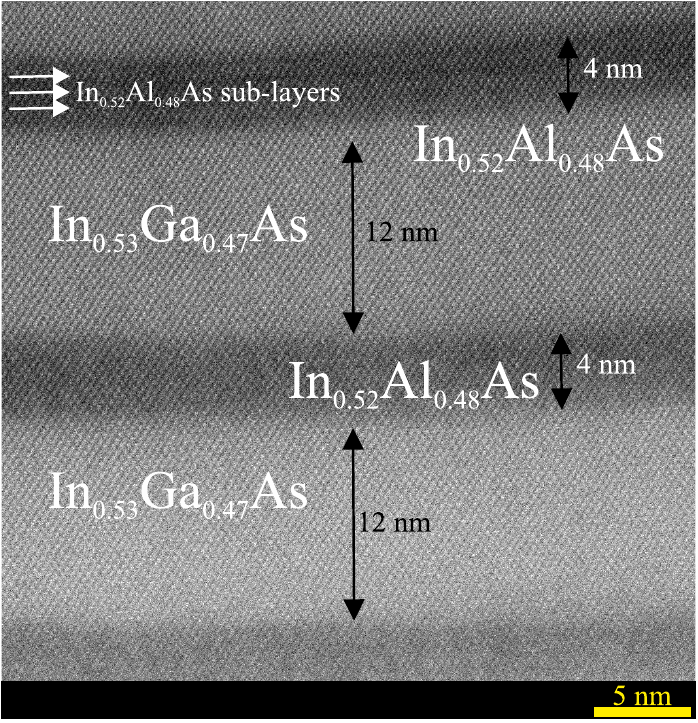}
    \caption{\small{Dark-field TEM-image of the non-strained  $\text{In}_{0.53}\text{Ga}_{0.47}\text{As/In}_{0.52}\text{Al}_{0.48}\text{As}$ superlattice with thicknesses of 12 nm and 4 nm, respectively. The $\text{In}_{0.52}\text{Al}_{0.48}\text{As}$ sub-layers, approximately 1$-$2 nm thick, referred to defect levels formation, are seen in the $\text{In}_{0.52}\text{Al}_{0.48}\text{As}$ barrier layers (indicated by horizontal arrows).}}
    \label{TEM}
\end{figure}

\subsection{Optical pump-probe and THz time-domain spectroscopy measurements}
\begin{figure}[h]
  \centering
    \includegraphics[width=0.9\columnwidth]{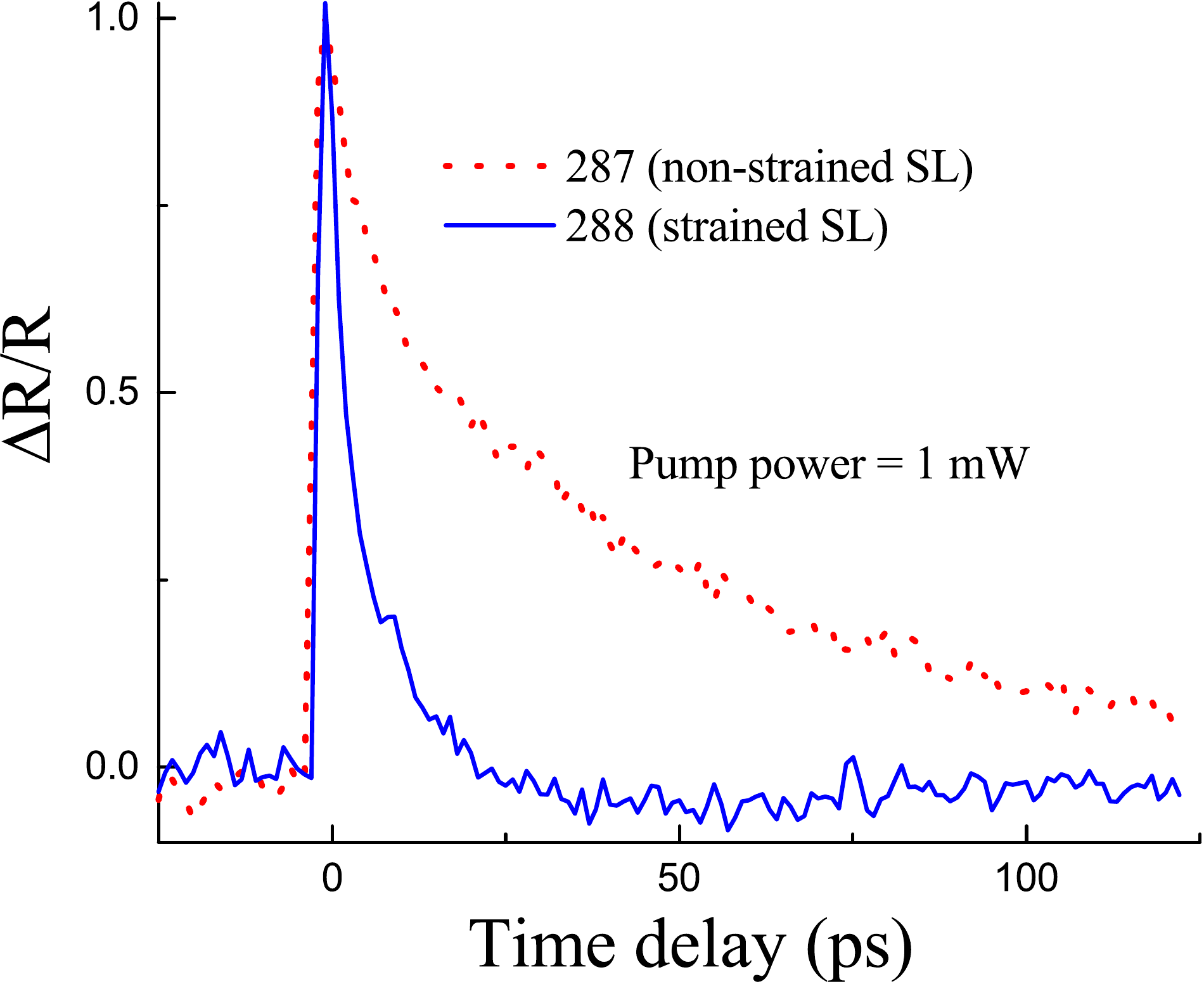}
    \caption{\small{Pump-probe measurements for the non-strained $\text{In}_{0.53}\text{Ga}_{0.47}\text{As/In}_{0.52}\text{Al}_{0.48}\text{As}$ (287) and
    strained $\text{In}_{0.53}\text{Ga}_{0.47}\text{As/In}_{0.38}\text{Al}_{0.62}\text{As}$ (288) superlattices at pump power of 1 mW (a) and 5 mw (b) and excitation wavelength of 800 nm.}}
    \label{TDS}
\end{figure}

Pump-probe transient reflectivity technique was used to measure photocarrier dynamics at different laser pumps. As excitation source, we used an ultrafast Ti:sapphire laser with central wavelength at 800 nm, 100 fs pulse duration, and 80 MHz repetition rate. Pump beam with an average power of 1 mW or 5 mW passed through delay line and was focused into the spot of $\sim 40\mu$m diameter on the sample surface. Probe beam with an average power of 40 mW was focused by 7.5 cm focal length lens on the sample surface and overlapped with pump beam. Reflected probe beam under 45$^{\circ}$ to normal was registered by the photodetector and lock-in amplifier. Polarization of the pump and probe pulses were crossed in order to avoid interference between pulses. Measurements were carried out at room temperature. For the time-zero point we chose the moment when two pulses simultaneously hit a surface of the sample.
Due to a complex nature of photocarrier dynamics in the SLs, for more accurate matching, the interpretation of the experimental results was made by three-exponential fit according to Ref.\cite{Kimel}:

\begin{equation}
\begin{gathered}
\frac{\Delta R}{R}(t_{d})=\sum_{i=1}^{3}\frac{A_i}{2}\exp\left(\frac{w^2}{4\tau_i^2}-\frac{\tau_d}{\tau_i}\right)\left[erf(\frac{\tau_d}{w}-\frac{w}{2\tau_i})+1\right].
\end{gathered}
\end{equation}
where $\Delta R$ is a change in probe reflectivity signal, generated by pump pulse; $R$ is signal of the probe reflectivity before pump hits the samples surface; $t_d$ is a time delay between two pulses; $A_{i}$ is the amplitude of the $\Delta R/R$; $w$ is a excitation time; $i$ is a number of the relaxation process; $\tau_{i}$ is the relaxation time corresponding to the different relaxation processes, respectively. Experimental transient reflectivity curves for all grown samples are depicted in Fig.~\ref{TDS}.

The observed relaxation times correspond to different processes occurring with photocarriers at various time intervals after photoexcitation, namely $\tau_{1,2}$ - are relaxation times attributed to hot charge carriers, at which the processes of intersubband scattering and scattering on acoustic phonons occur, $\tau_{3}$ - is a conduction-to-valence band recombination time. We considered the first two relaxation times, since they correspond to the loss of energy of photocarriers and its capture by the recombination centers, and thus, they are of particular importance when fabricating SL-based photoconductive antennas. In this case, the behavior of photocarriers and transient processes immediately after photoexcitation, such as thermalization and scattering on optical phonons, are not visible on the long interval of the used time delays. The values of the relaxation times $\tau_{1,2}$ for the two pump powers are given in Table.1. It should be noticed that we limited pump-probe measurements by the pump power of 5 mW, since higher powers lead to traps saturation, and hence, the obtained relaxation times for the SLs will significantly increase  \cite{DietzOPTEXP}.

\begin{table}[!b]
\centering
 \caption{Relaxation times $\tau_{1,2}$ extracted from the pump-probe measurements for the samples under consideration for two pump powers and excitation wavelength of 800 nm.}
  \includegraphics[width=1.0\columnwidth]{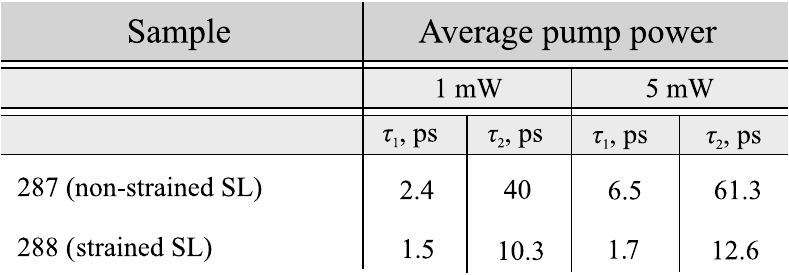}
  \label{TAB:Table1}
\end{table}

As it can be seen from the Table~\ref{TAB:Table1} and Fig~\ref{TDS}, at both pump powers of \mbox{1 mW} and \mbox{5 mW} the smallest relaxation times \mbox{$\tau_{1}$=1.7 ps} and \mbox{$\tau_{1}$=1.5 ps} are obtained for the sample 288. We assume that this is attributed not only to the existence of defect levels in the $\text{In}_{0.52}\text{Al}_{0.48}\text{As}$ barriers that effectively provide non-radiative recombination centers but also to the created stresses in the barriers which result in scattering of photocarriers at the $\text{In}_{0.53}\text{Ga}_{0.47}\text{As/In}_{0.38}\text{Al}_{0.62}\text{As}$ heterointerface roughness \cite{Gottinger,Sakaki}, thus, significantly decreasing the relaxation times. Increasing the pump power up to \mbox{5 mW} leads to saturation of relaxation times in both samples due to high density of photocarriers, which is comparable and even higher than a density of recombination centers. In this case photocarriers are filling all vacant states at defect levels, significantly increasing relaxation times \cite{DietzOPTEXP}.

To measure the THz pulse amplitude of the samples under investigation we used the THz time-domain spectroscopic (THz-TDS) setup.
Pump beam with an
average power of 85 mW after passage through
the delay line was focused on the samples surface by an
objective with fivefold de-magnification. Probe beam had an average power of 40 mW. The generated THz pulse was collected by the first parabolic mirror and
then focused on the surface of ZnTe detector by the second parabolic mirror. A
detailed scheme of the setup is presented in Ref.\cite{Khusyainov2017}. The
probe pulse transmitted through ZnTe
was detected by a photodiode and a synchronous
detector. On the path of the probe pulse, a polarizer
and an analyzer with crossed polarizations were
placed.

\begin{figure}[h]
  \centering
    \includegraphics[width=1.0\columnwidth]{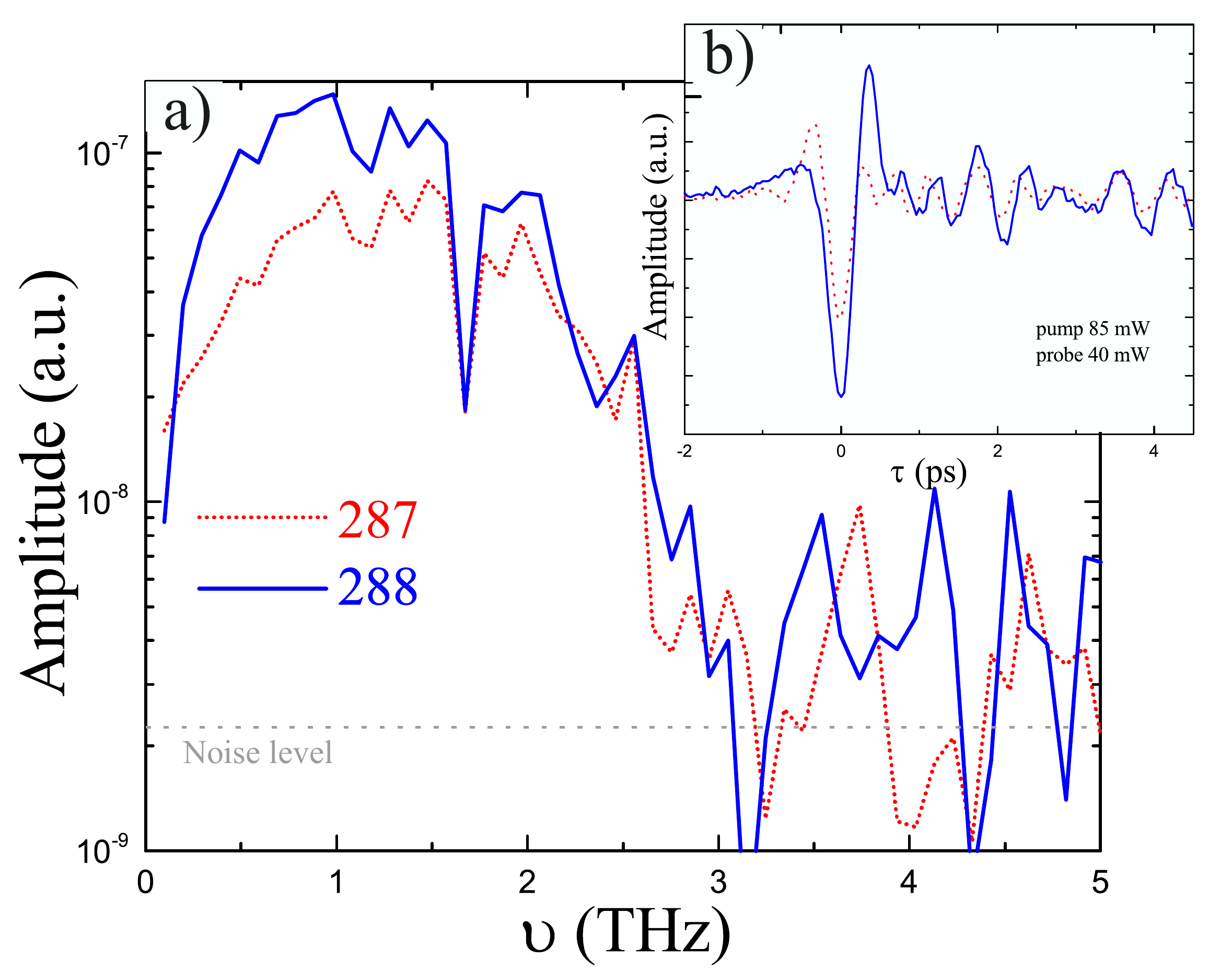}
    \caption{\small{The amplitude of the emitted THz power spectra (a) and the waveforms of THz pulses for the non-strained $\text{In}_{0.53}\text{Ga}_{0.47}\text{As/In}_{0.52}\text{Al}_{0.48}\text{As}$ (287) and strained $\text{In}_{0.53}\text{Ga}_{0.47}\text{As/In}_{0.38}\text{Al}_{0.62}\text{As}$ (288) superlattices at pump power of 85 mW and excitation wavelength of 800 nm.}}
    \label{TDS2}
\end{figure}

The results of TDS measurements are demonstrated in Fig.~\ref{TDS2}. The resonant frequency is equal to 1 THz for all the samples under investigation, while the notch close to 1.67 THz is associated with the
resonant THz absorption by water vapour along the beam
path; however, it does not prevent analysis and
comparison of the emitted THz power spectra of the samples.
As it can be seen from Fig.~\ref{TDS2}~(b) an amplitude of a waveform of THz pulse approximately twice increases when comparing the non-strained (287) and the strained (288) SLs. As a result, in (a) one can see that the emitted THz power spectra is increasing and slightly broadening with shift into the high frequency region \cite{DietzOPTEXP} with residual strain increase. We believe the two dominant mechanisms being responsible for the sharp reduction of relaxation time and consequent increase of THz power spectra that are photocarriers trapping by the defect levels in the $\text{In}_{\text y}\text{Al}_{1-\text y}\text{As}$ (where $\text y$ is the indium mole fraction) barriers and increased photocarriers scattering at the $\text{In}_{0.53}\text{Ga}_{0.47}\text{As/In}_{0.38}\text{Al}_{0.62}\text{As}$ heterointerface roughness.

\section{Conclusions}

We have investigated an influence of residual strain in InGaAs/InAlAs superlattice (SL) on the ultrafast photocarrier dynamics under femtosecond laser excitation. We proposed and fabricated a novel-designed strained InGaAs/InAlAs SL, which allows for obtaining an ultrashort photocarrier relaxation time $\tau\sim$~1.5~ps without Be-doping of InGaAs photoconductor. We assumed two dominant mechanisms being responsible for a sharp reduction of $\tau$ which are: photocarrier trapping by the defect levels in InAlAs barriers and increased photocarrier scattering at InGaAs/InAlAs interface roughness due to residual strain in the SL. The THz time-domain spectroscopy measurements revealed an increase of the amplitude in both THz pulse waveform and, consequently, THz emitted spectrum with an increase of the residual strain in SL. The achieved results will be consequently used for fabrication of SL-based photoconductive antennas with enhanced performance, particularly under longer-wavelength (up to 1.55~$\mu$m) laser pump.
\section{Acknowledgments}
The authors are grateful to A.Klochkov for the help with numerical simulations and to A.Vasil'ev for TEM measurements.
The work was supported by the Russian Scientific Foundation (Project No.18-79-10195), the Russian Foundation for Basic Research, grants Nos. 18-02-00843 and 16-29-14029, and RIEC ICRP
(grant H30/A04). The epitaxial growth of the samples under consideration was supported by the President's grant of Russian Federation for young scholars, Project MK-5450.2018.2.


\end{document}